\documentclass[10pt,aps,prl,preprintnumbers,superscriptaddress,showpacs,
twocolumn,floatfix,nofootinbib]{revtex4-1}
\usepackage{graphicx}
\usepackage{amsmath}
\usepackage{slashed}
\usepackage{hyperref}

\newcommand{\gsim}{\;\rlap{\lower 3.5 pt \hbox{$\mathchar \sim$}} \raise 1pt
\hbox {$>$}\;}
\newcommand{\lsim}{\;\rlap{\lower 3.5 pt \hbox{$\mathchar \sim$}} \raise 1pt
  \hbox {$<$}\;}
\newcommand{\be}{\begin{equation}}
\newcommand{\ee}{\end{equation}}

\begin{document}

\title{\boldmath
Mixed QCD-electroweak corrections to on-shell $Z$ production at the LHC
\unboldmath}
\author{Federico Buccioni}
\email[]{federico.buccioni@physics.ox.ac.uk}
\affiliation{Rudolf Peierls Centre for Theoretical Physics, Clarendon Laboratory,
  Parks Road, Oxford OX1 3PU, UK}

\author{Fabrizio Caola}
\email[]{fabrizio.caola@physics.ox.ac.uk}
\affiliation{Rudolf Peierls Centre for Theoretical Physics, Clarendon Laboratory,
  Parks Road, Oxford OX1 3PU, UK}
\affiliation{Wadham College, University of Oxford, Parks Rd, Oxford OX1 3PN, UK}

\author{Maximilian Delto }
\email[]{maximilian.delto@kit.edu}
\affiliation{Institut f\"ur Theoretische Teilchenphysik,
 Karlsruher Institut f\"ur Technologie (KIT), 76128 Karlsruhe, Germany}

\author{Matthieu~Jaquier}
\email[]{matthieu.jaquier@kit.edu}
\affiliation{Institut f\"ur Theoretische Teilchenphysik,
 Karlsruher Institut f\"ur Technologie (KIT), 76128 Karlsruhe, Germany}

\author{Kirill Melnikov}
\email[]{kirill.melnikov@kit.edu}
\affiliation{Institut f\"ur Theoretische Teilchenphysik,
  Karlsruher Institut f\"ur Technologie (KIT), 76128 Karlsruhe, Germany}

\author{Raoul R\"ontsch}
\email[]{raoul.rontsch@cern.ch}
\affiliation{Theoretical Physics Department, CERN, 1211 Geneva 23, Switzerland}

\begin{abstract}
          We present the first complete calculation of mixed QCD-electroweak corrections to the production of on-shell 
      $Z$ bosons in hadron collisions and their decays to massless charged leptons.
      Our computation is fully differential with respect to final state
      QCD partons and resolved  photons, allowing us to compute any infra-red safe observable pertinent to the $pp \to Z \to l^+ l^-$
      process in the approximation that the $Z$ boson is on shell. 
      Although mixed QCD-electroweak  corrections are small, at about the per mill level,   we observe
      that the interplay between QCD-QED  and QCD-weak contributions is subtle and observable-dependent.  It is therefore
     not possible to avoid computing one or the other if ${\cal O}(\alpha_{EW} \alpha_s)$ precision is desired. 

\end{abstract}
\preprint{OUTP-20-04P,TTP20-021, P3H-20-019, CERN-TH-2020-072}

\maketitle

    The production of lepton pairs in hadron collisions $pp \to l^+
    l^-$ has played and continues to play an important role in the
    exploration of the inner workings of the Standard Model (SM) and
    in ongoing attempts to access physics beyond it.  A seminal 1970
    paper by Drell and Yan \cite{dy} pointed out the connection between a
    theoretical description of this process and the parton model of
    deep-inelastic scattering. This observation initiated the
    development of the quantitative theory of lepton pair production
    in hadron collisions~\cite{Altarelli:1978id,Altarelli:1979ub,
      Parisi:1979se} and
    encouraged its experimental exploration~\cite{Christenson:1970um,
    Christenson:1973mf}.  Subsequent
      theoretical developments in perturbative QCD and in the SM
      resulted in a continuously improving description of this process
      and provided a solid foundation for ambitious experimental
      studies aiming at measuring the SM parameters with high
      precision and at constraining New Physics.
    
    Indeed, the production of lepton pairs at the LHC is the process
    from which the mass of the $W$ boson is expected to be determined with
    an astounding precision of about $5~{\rm
      MeV}$~\cite{Aaboud:2017svj,ATLAS:2018qzr}.  This process is also
    very important for constraining parton distribution
    functions~\cite{Khachatryan:2016pev,Aaboud:2016btc} and for
    determining the electroweak
    mixing angle~\cite{Aad:2015uau,Chatrchyan:2011ya}. Finally, it can
    be used to constrain higher-dimensional operators which
    parametrize deviations from the SM by studying the invariant mass
    distribution of the dilepton system at high ${\cal O}(1~{\rm
      TeV})$ invariant masses~\cite{Khachatryan:2016zqb,
      Khachatryan:2016jww,Aaboud:2016cth,Aaboud:2016zkn}. An obvious
    pre-requisite for the success of this challenging research program
    is the existence of a reliable theoretical description of all
    aspects of lepton pair production in hadron collisions.

    A central role in providing such a description is played by
    perturbative calculations in the Standard Model.  Currently, the
    fully-differential cross sections for dilepton production in
    hadron collisions are known through next-to-next-to-leading order
    (NNLO) in perturbative QCD~\cite{Melnikov:2006di,Melnikov:2006kv,
      Catani:2009sm,Catani:2010en,Boughezal:2016wmq,Caola:2019nzf,
      Gavin:2010az,Li:2012wna,Gavin:2012sy,Grazzini:2017mhc,Camarda:2019zyx}
    and through next-to-leading order (NLO) in the electroweak
    theory~\cite{Dittmaier:2001ay,
      Baur:2004ig,Zykunov:2006yb,Arbuzov:2005dd,CarloniCalame:2006zq,
      Baur:2001ze,Zykunov:2005tc,CarloniCalame:2007cd,Arbuzov:2007db,
      Dittmaier:2009cr}. Recently the inclusive cross section of the process
    $pp \to \gamma^* \to l^+ l^-$ has been computed through N$^3$LO in
    perturbative QCD \cite{claude}. Important steps in further
    increasing precision are the extension of this N$^3$LO result to
    the case of $Z$ and $W$ production and the calculation of the
    so-called mixed QCD-electroweak ${\cal O}(\alpha_{EW} \alpha_s)$
    corrections.  The latter class of corrections is the subject of
    the present paper.
 
    The computation of mixed QCD-electroweak corrections is made
    complicated by the fact that they require broad technical
    expertise. Indeed, on the one hand, one has to compute two-loop
    three- and even four-point functions with massive internal and
    external particles yet, on the other hand, a detailed
    understanding of infra-red and collinear singularities and their
    regularization is also needed.

    There is quite a number of different physical
    aspects of dilepton pair production in hadron collisions that gets
    reflected in technical complexities of theoretical computations.
    This implies that by choosing a suitable {\it physical
      problem}, one may scale the technical complexity up or
    down.  Indeed, if one focuses on QCD-electroweak corrections to
    the full production of dileptons at e.g.  high invariant masses,
    two-loop virtual corrections involve box diagrams that depend on
    several mass scales. The corresponding master integrals have been
    computed recently \cite{Bonciani:2016ypc,
      Heller:2019gkq,Hasan:2020vwn} and the scattering amplitudes are
    still not available.  On the
    contrary, if one focuses on on-shell $Z$ or $W$ production,
    the cross-talk between production and decay stages of the process
    is suppressed\footnote{A concise discussion of physical reasons
      behind this suppression can be found in
      Ref.~\cite{Fadin:1993kt}.}~\cite{dit1,dit2} so that the most
    complicated two-loop contributions one has to consider are
    two-loop corrections to the $q \bar q' \to Z(W)$ vertex.  Similarly,
    if one considers the production of on-shell $Z$ bosons,
    the regularization of infra-red and collinear singularities in mixed
    contributions simplifies since NNLO-like emissions of a photon and
    a gluon can only happen in the production stage. In this
    sense, the treatment of infra-red and collinear singularities is
    very similar to what happens when computing NNLO QCD
    corrections to $Z$ boson production.

    Thanks to these significant technical simplifications, it is quite
    natural that physical results for mixed QCD-electroweak corrections
    to $pp \to l^+l^-$ started to appear in the context of on-shell
    $Z$ boson production.  In Ref.~\cite{deflorian1} it was pointed out that
    a simple modification of color factors in an analytic result for
    NNLO QCD corrections to the total cross section of $pp \to l^+
    l^-$ \cite{ham1,ham2,har} allows one to obtain mixed QCD-QED
    corrections to the total cross section of dilepton production.  In
    Ref.~\cite{Delto:2019ewv} some of us performed a
    fully-differential computation of these QCD-QED corrections to $Z$
    production and decay into a pair of massless leptons adapting the
    soft-collinear subtraction scheme~\cite{Caola:2017dug} developed
    for NNLO QCD computations to describe mixed QCD-QED
    effects. Similar calculations were reported  in
    Refs.~\cite{Buonocore:2019puv,Cieri:2020ikq} within the $q_T$
    slicing framework. 
    
    The next natural step is to extend these results to include mixed
    QCD-{\it weak} corrections to the description of on-shell
    $Z$ production and its subsequent decay to a pair of massless
    electrons. A first step in this direction was done
    in~\cite{Bonciani:2019nuy}.
    The main of focus of the current paper is to
    consider the process $pp\to Z \to e^+ e^- +X$ at full $\mathcal
    O(\alpha_{EW}\alpha_s)$, in the approximation that the $Z$ boson is
    on shell and electrons are massless.  From the phenomenological
    point of view, the knowledge of mixed corrections to on-shell
    $Z$ production is perhaps not extremely interesting but the
    calculation of these corrections is a good starting point for the
    analysis of the much more interesting case of the $W$ boson
    production.

      Since, by definition, weak corrections include exchanges of
      massive gauge bosons, mixed QCD-weak corrections do not contribute to
      genuine NNLO infra-red and collinear divergences; all such
      divergences reside in mixed QCD-QED corrections which have
      already been studied in Ref.~\cite{Delto:2019ewv}.  Hence, from
      a technical point of view, the inclusion of mixed QCD-weak
      corrections requires the computation of one- and two-loop mixed
      QCD-weak contributions to e.g. $q \bar q
      \to Z+g$ and $q \bar q \to Z$ amplitudes as well as
      their renormalization.

\begin{table}[t]
\begin{center}
\begin{tabular}{|c|c|c|c|}
\hline \hline
Type   & Inclusive & Cuts  & Cuts (production) \\
\hline 
$ \Delta_{\rm NLO}^{\rm QED}$  & $ +2.3 \times 10^{-3} $  & $-5.3 \times 10^{-3}$ &  $+2.2 \times 10^{-3}$  \\
$ \Delta_{\rm NLO}^{\rm weak}$ &  $ -5.5 \times 10^{-3} $ &  $-5.0 \times 10^{-3}$ & $-5.0 \times 10^{-3}$ \\
$ \Delta_{\rm NLO}^{\rm EW}$ &  $ -3.2 \times 10^{-3} $ &  $-1.0 \times 10^{-2}$ & $-2.8 \times 10^{-3}$ \\
\hline 
$ \Delta_{\rm NNLO}^{\rm QCD-QCD}$ & $+1.3 \times 10^{-2} $     &   $+5.8 \times 10^{-3}$ & $+5.8 \times 10^{-3} $\\
\hline 
$ \Delta_{\rm NNLO}^{\rm QCD-QED}$ & $+5.5 \times 10^{-4} $      &  $-5.9 \times 10^{-3} $ & $ +1.4 \times 10^{-4}$ \\
$ \Delta_{\rm NNLO}^{\rm QCD-weak}$  & $-1.6 \times 10^{-3}$   &   $ -2.1 \times 10^{-3} $ & $ -2.1 \times 10^{-3}$ \\ 
$ \Delta_{\rm NNLO}^{\rm QCD-EW} $ & $-1.1 \times 10^{-3}$      &   $ -8.0 \times 10^{-3} $ &  $-2.0 \times 10^{-3}$\\
\hline \hline 
\end{tabular}
\end{center}
\caption{
  Corrections to the total cross section of $pp \to Z \to e^+ e^-$ in the narrow width approximation at the $13~{\rm TeV}$ LHC.
See text for further details.}
\end{table}

  We computed two-loop QCD-electroweak corrections to the $q \bar q
  \to Z$ vertex using standard techniques and found agreement with
  available results in the
  literature~\cite{Kotikov:2007vr}.\footnote{We note that we do not
    include the finite part of 
    two-loop contributions involving
    exchanges of virtual top quarks.  At one-loop these contributions
    amount to about $10$ percent of the full one-loop weak virtual
    correction.}  We extracted the ingredients required for the two-loop
  mixed QCD-electroweak renormalization from
  Ref.~\cite{Djouadi:1993ss}.\footnote{ We note that there is a small typo
    in Eq.~(5.4) of this reference.}  We obtained one-loop weak
  corrections to $q \bar q \to Z+g$ and related partonic channels
  numerically using the OpenLoops package
  \cite{Cascioli:2011va,Buccioni:2017yxi,Buccioni:2019sur}. In OpenLoops scalar integrals
  are provided by~\cite{Denner:2016kdg,vanHameren:2010cp}.
  The renormalization of weak corrections is
  performed in the $G_\mu$ scheme;\footnote{See
    e.g. Ref.~\cite{Denner:2019vbn} for a review.} the strong
  coupling constant is renormalized in the $\overline{\rm MS}$
  scheme. Numerically, we use $G_F = 1.16639 \times 10^{-5}~{\rm
    GeV}^{-2}$, $M_Z = 91.1876~{\rm GeV}$, $M_W = 80.398~{\rm GeV}$,
  $M_{t} = 173.2~{\rm GeV}$ and $M_H=125~{\rm GeV}$ as input
  parameters. With this setup, we obtain $1/\alpha = 132.338$ for the
  fine-structure constant.
  We use the NNLO NNPDF3.1luxQED~\cite{Bertone:2017bme,
    Manohar:2017eqh,Manohar:2016nzj} parton distribution functions for
  \emph{all} numerical computations.  The value of the strong coupling constant
  is provided as part of the PDF set; numerically it reads
  $\alpha_s(M_Z) = 0.118$.
     
     We also employ standard kinematic selection criteria by requiring that
     the transverse momenta of the two leptons satisfy $p_{t,l} >
     24(16)~{\rm GeV}$ for the harder(softer) lepton.
     The rapidities of the two leptons should satisfy
     $-2.4 < y_{l} < 2.4 $.  Finally, since we neglect lepton
     masses, we have to {\it define} photons and leptons in a way that
     is robust against the collinear splittings $e \to e+\gamma$.  To this
     end, leptons and photons are clustered into ``lepton jets''
     provided that the angular distance $ R_{e\gamma} = \sqrt{ (y_e
       - y_\gamma)^2 + (\varphi_e - \varphi_\gamma)^2} $ between $e$ and
     $\gamma$ is smaller than $0.1$~\cite{simone}. The reconstructed
     dilepton system is required to have an invariant mass greater
     than $50~{\rm GeV}$.  For all results reported below, we choose
     the renormalization scale of the strong coupling constant and the
     factorization scale in parton distributions to be $\mu_R = \mu_F
     = M_Z/2$.

    We compute the production of the $Z$ boson in the narrow width
    approximation.  We find it convenient to re-write the differential
    cross section as
    \be {\rm d} \sigma_{pp \to e^+e^-} = {\rm Br}(Z
    \to e^+ e^-) \; {\rm d} \sigma_{pp \to Z} \frac{{\rm d} \Gamma_{Z
        \to e^+ e^-} }{\Gamma_{Z \to e^+ e^-} },
       \label{eq1}
    \ee
    factoring out the branching fraction ${\rm Br}(Z \to
    e^+e^-)$. We {\it do not} perform a perturbative expansion of
    ${\rm Br}(Z \to e^+ e^-)$.
    In what follows, we will consider ratios of cross sections and
    kinematic distributions, so the branching ratio drops from our
    results.\footnote{We note that mixed QCD-electroweak corrections to
      ${\rm Br}(Z \to e^+ e^-) $ can be extracted from
      Ref.~\cite{Czarnecki:1996ei}.}  All other contributions in
    Eq.~(\ref{eq1}) are expanded in powers of $\alpha_{EW}$ and
    $\alpha_s$. For further details, the reader should consult
    Ref.~\cite{Delto:2019ewv}.

\begin{figure*}[t]
  \centering
  \begin{minipage}[t]{0.48\textwidth}
    \includegraphics[clip,width=1\textwidth,page=1,angle=0]{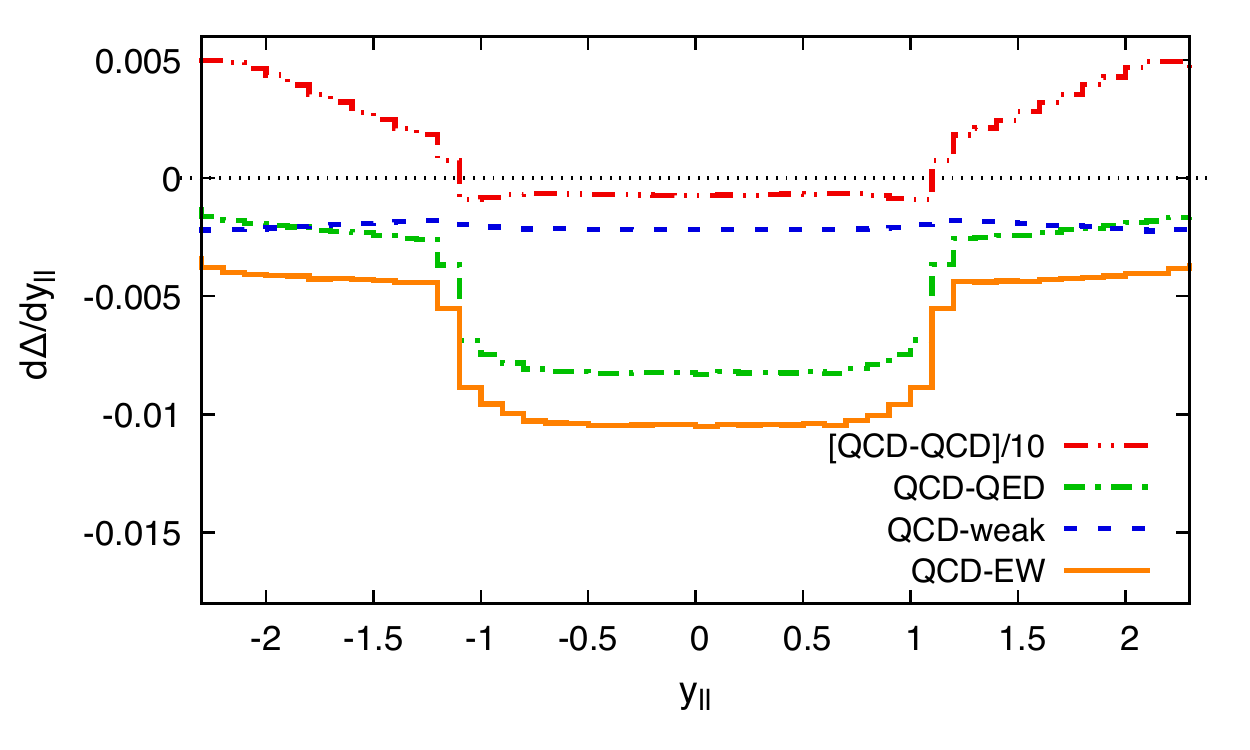}
  \end{minipage}
  \hfill
  \begin{minipage}[t]{0.48\textwidth}
    \includegraphics[clip,width=1\textwidth,page=1,angle=0]{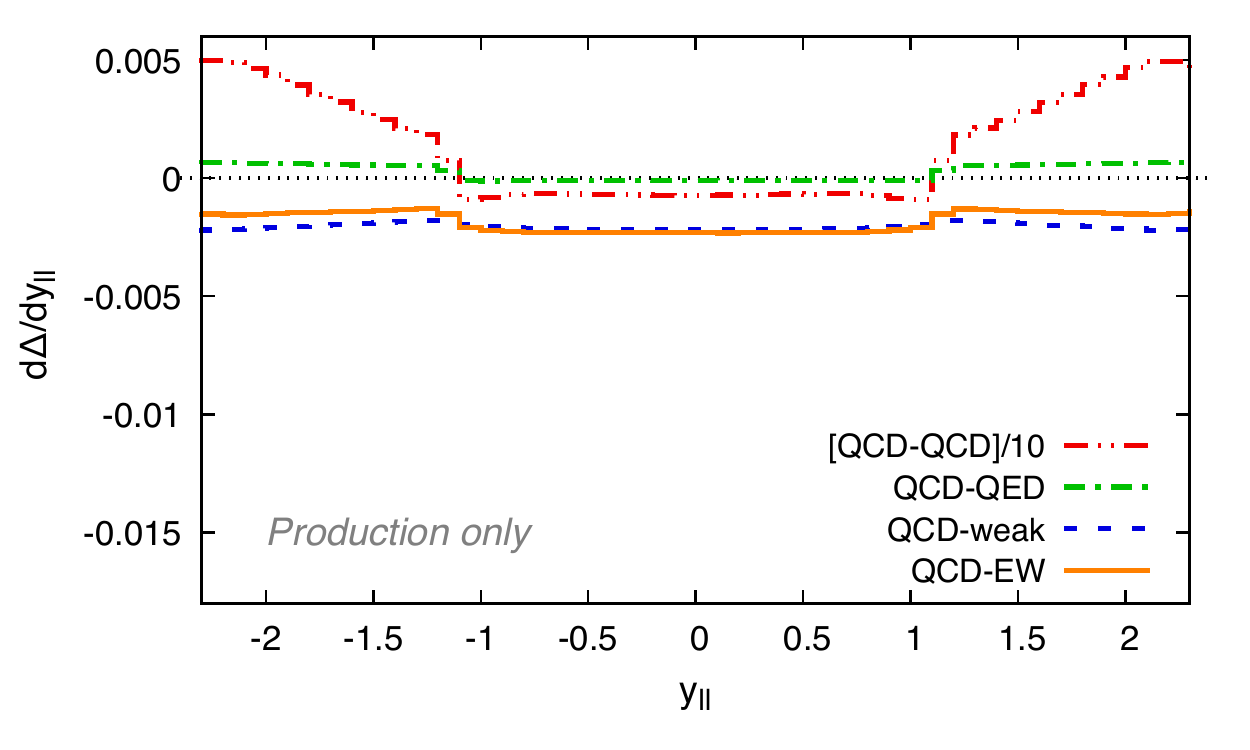}
  \end{minipage}
  \\
  \begin{minipage}[t]{0.48\textwidth}
    \includegraphics[clip,width=1\textwidth,page=1,angle=0]{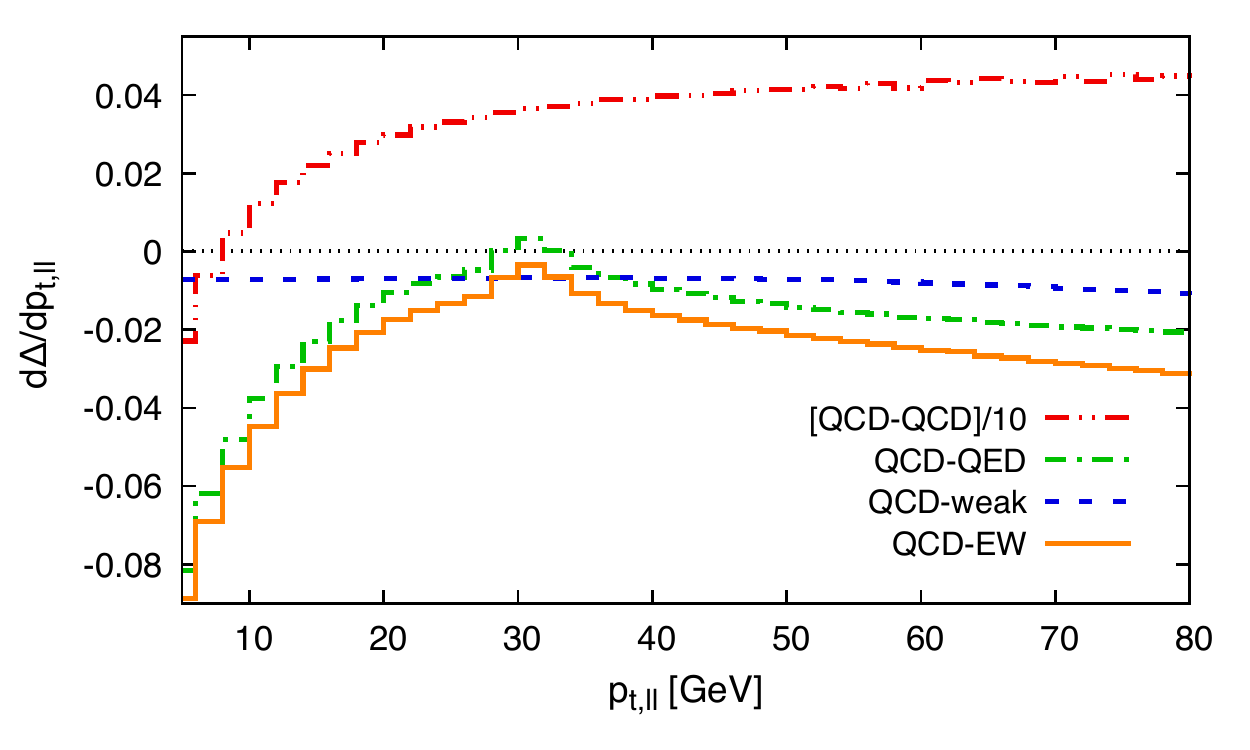}
  \end{minipage}
  \hfill
  \begin{minipage}[t]{0.48\textwidth}
    \includegraphics[clip,width=1\textwidth,page=1,angle=0]{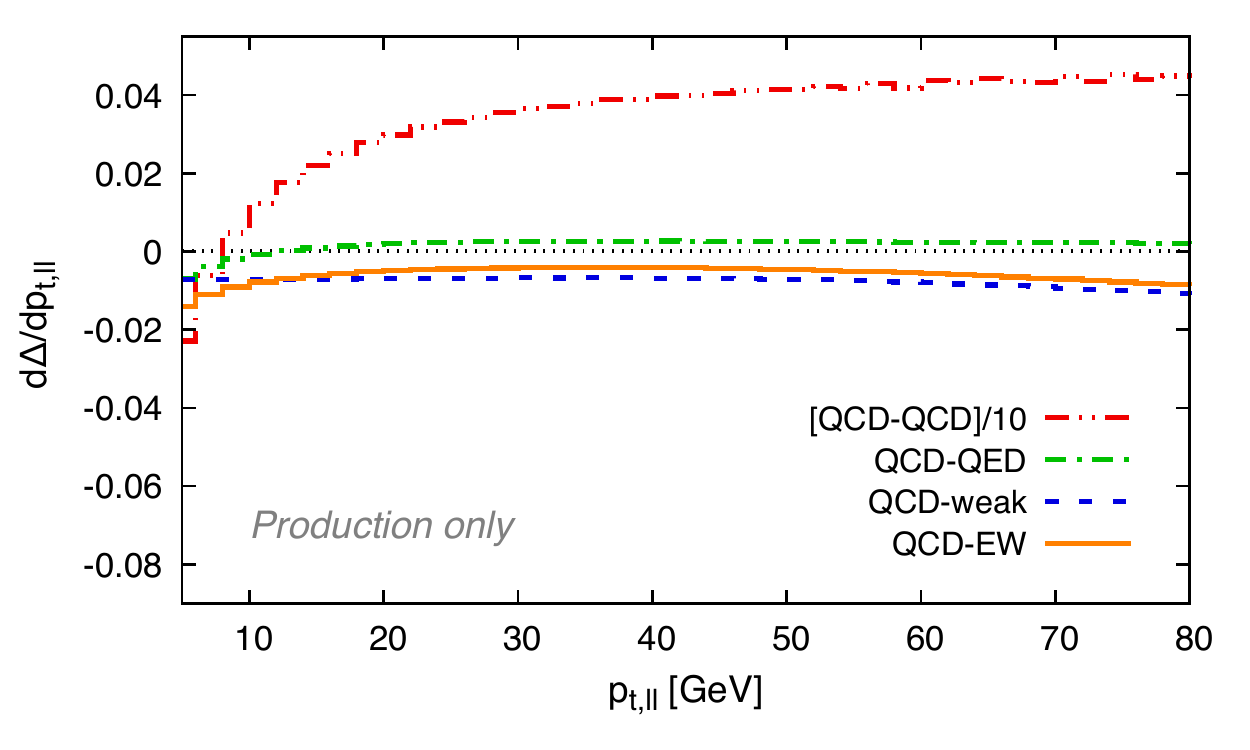}
  \end{minipage}
 \caption{Mixed QCD-electroweak corrections to dilepton rapidity and
   transverse momentum distributions at the $13~{\rm TeV}$ LHC.  Left
   pane includes corrections to both production and decay whereas
   right pane includes corrections to the production stage only.  See text
   for details. }
\end{figure*}

To present our results, we expand     the cross section of the process
$pp \to Z \to e^+e^-$   in series in  $\alpha_s$ and $\alpha_{EW}$
\be
\begin{split} 
{\rm d} \sigma  = {\rm d} \sigma_{\rm LO} & + {\rm d}  \sigma_{\rm NLO}^{\rm QCD} +  {\rm d} \sigma_{\rm NLO}^{\rm EW} 
\\
&
+ {\rm d} \sigma_{\rm NNLO}^{\rm QCD-QCD}
+ {\rm d}\sigma_{\rm NNLO}^{\rm QCD-EW}+...
\end{split} 
\ee
The new result that we describe in this paper is the mixed QCD-electroweak
contribution ${\rm d}\sigma_{\rm NNLO}^{\rm QCD-EW}$.  This
contribution is the sum of QCD-QED and QCD-weak
corrections; in what follows we will show these 
contributions separately.
    
We find it convenient to quote ratios of NLO electroweak and NNLO
contributions to the NLO QCD differential
cross section.
Hence, we define
\be
{\rm d} \Delta^{i} = \frac{{\rm d} \sigma^{i} }{ {\rm d}
  \sigma_{\rm LO} + {\rm d} \sigma^{\rm QCD}_{ \rm NLO} },
\ee
where $i\in\{{\rm EW},{\rm QCD-EW},{\rm QCD-QCD}\}$, and ${\rm EW}$
can be further split in ${\rm QED}$ and ${weak}$.
Furthermore, we will also show corrections to the production stage by
themselves.

We begin by discussing corrections to the total (inclusive) cross
section where no restrictions on the kinematics of the final state particles
are applied.  The corresponding results for the 13 TeV LHC are shown in
the second column of Table~1. We observe that NNLO QCD corrections
exceed mixed QCD-electroweak ones by almost one order of
magnitude. Interestingly, mixed corrections are dominated by weak
ones; they are larger than mixed QCD-QED corrections by almost a
factor of three.  Moreover, there is a cancellation between QCD-QED and
QCD-weak corrections so that the combined QCD-electroweak effect is about
one permille. Note that since we factorize the branching ratio ${\rm
  Br}(Z \to e^+ e^-)$, corrections to the decay have no bearing on the
inclusive cross section so that results in Table~1 can be regarded as
corrections to the inclusive process $pp \to Z$.
We note that mixed QCD-electroweak corrections do not make an appreciable
change to the scale uncertainty, which is still dominated by NNLO QCD
contributions. For this reason, an assessment of how the present calculation
reduces theoretical uncertainties on the $Z$ boson production cross section
will strongly depend on the quality of available QCD predictions.
Hence, the completion of N$^3$LO QCD calculations for this class of
processes becomes even more relevant.

\begin{figure*}
  \centering
  \begin{minipage}[t]{0.48\textwidth}
    \includegraphics[clip,width=1\textwidth,page=1,angle=0]{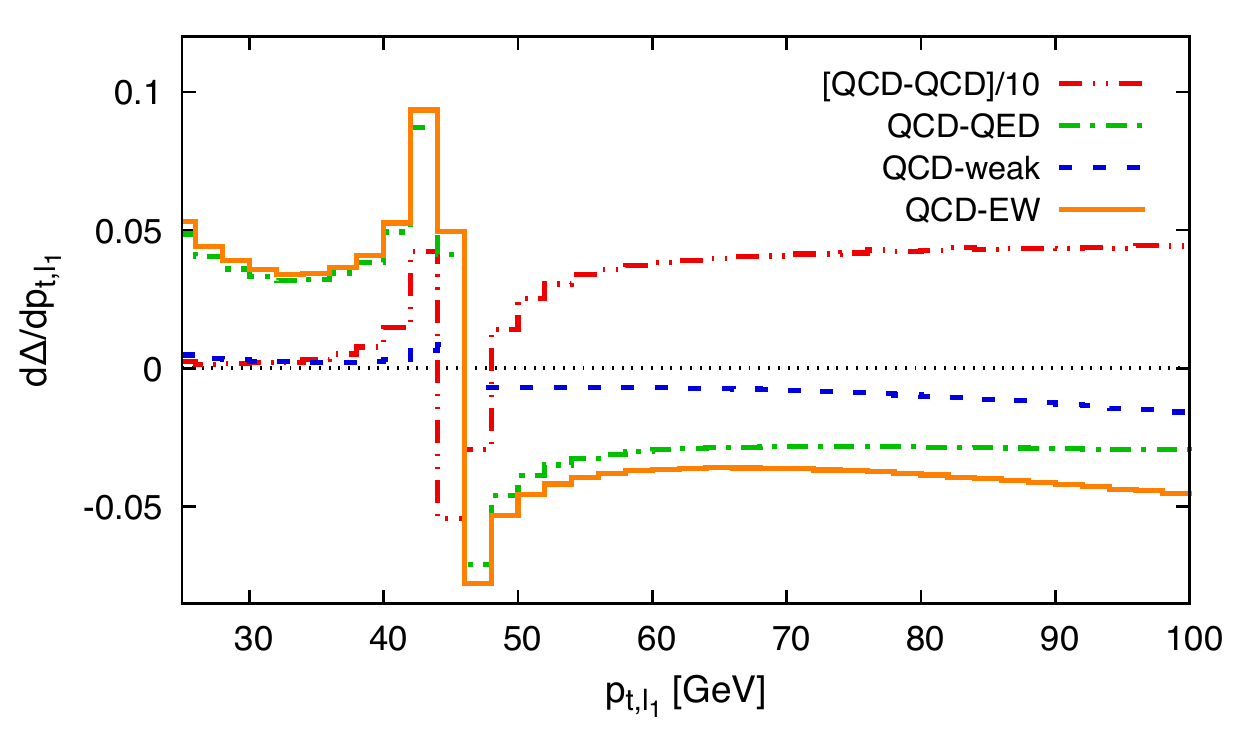}
  \end{minipage}
  \hfill
  \begin{minipage}[t]{0.48\textwidth}
    \includegraphics[clip,width=1\textwidth,page=1,angle=0]{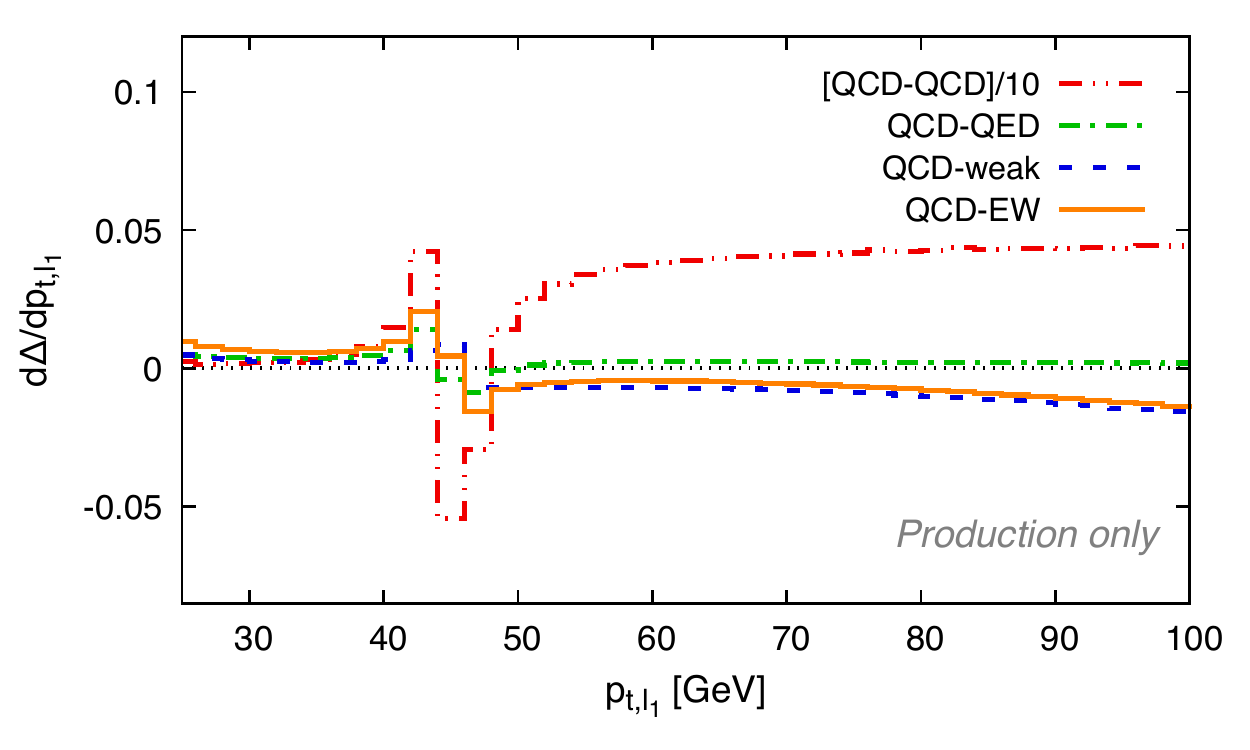}
  \end{minipage}
  \\
  \begin{minipage}[t]{0.48\textwidth}
    \includegraphics[clip,width=1\textwidth,page=1,angle=0]{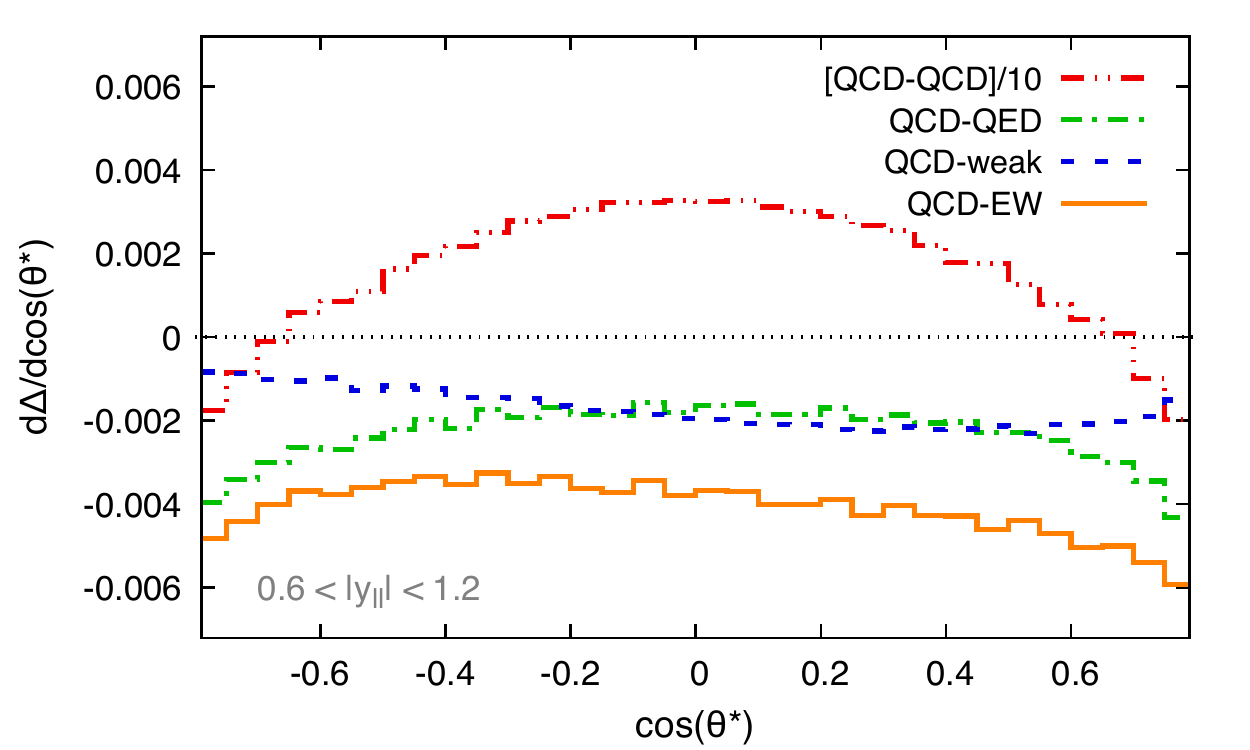}
  \end{minipage}
  \hfill
  \begin{minipage}[t]{0.48\textwidth}
    \includegraphics[clip,width=1\textwidth,page=1,angle=0]{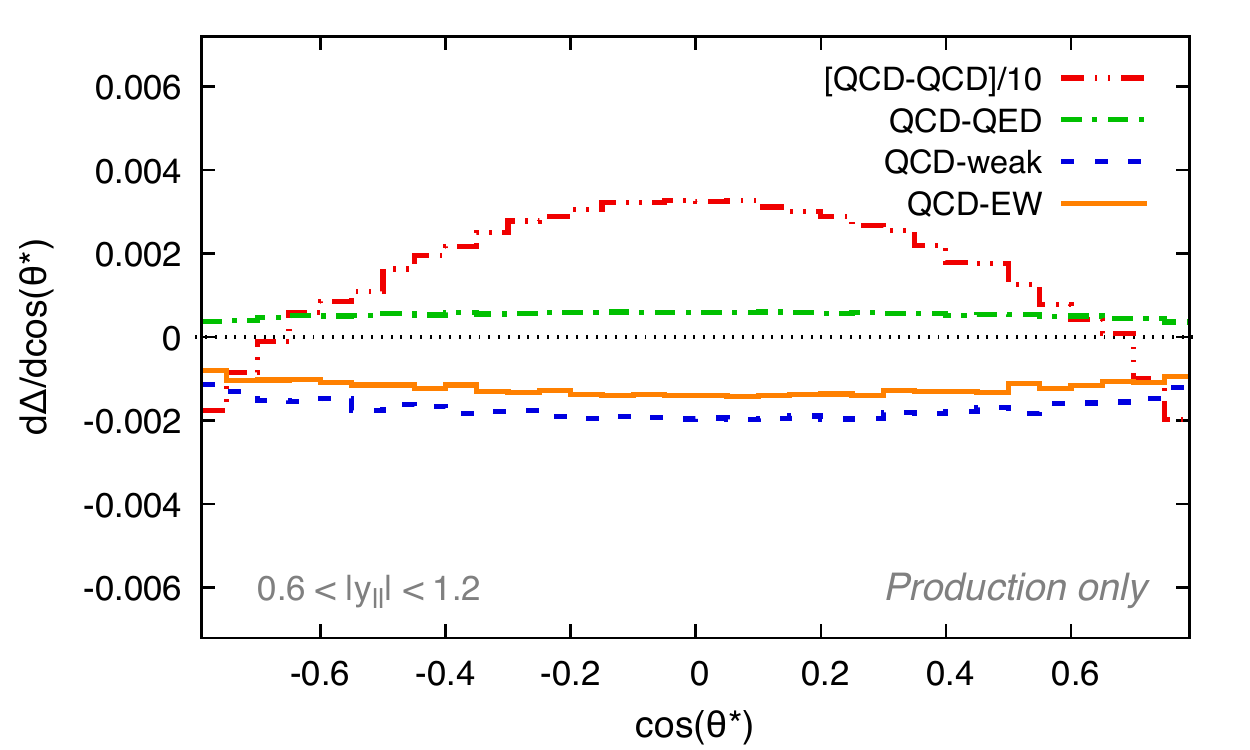}
  \end{minipage}
\caption{ Mixed QCD-electroweak corrections to dilepton rapidity and
  transverse momentum distributions at the $13~{\rm TeV}$ LHC.  Left
  pane includes corrections to both production and decay whereas right
  pane includes corrections to the production stage only.  See text for
  details.}
\end{figure*}

    The results change significantly when cuts to final state leptons
    are applied, see third column in Table~1. First, NNLO QCD corrections  decrease so strongly that mixed QCD-electroweak
    contributions become very relevant. 
    This is the consequence of
    an accidental cancellation between $q \bar q$ and $q g$ channels that appears to be quite dramatic once fiducial cuts
    are applied and the renormalization and factorization scales $\mu = M_Z/2$ are chosen. 
    Among electroweak corrections, the change   mostly concerns 
    QED corrections  which flip  sign relative to the
    inclusive case and increase by an order of magnitude.
    The latter issue  is well-known since  QED corrections to  $Z$ decays
    appear to be quite unstable for the set of fiducial cuts defined earlier.
    However, to the best of our knowledge a thorough study of how to ameliorate
    this situation has not been done yet. 
    The change in sign of
    the QED corrections implies that instead of a cancellation between QED
    and weak contributions occurring in the inclusive cross section,
    they add up in the case of the fiducial one. As the consequence, the
    QCD-electroweak corrections {\it exceed} 
    the NNLO QCD corrections in this case.\footnote{We emphasize again that this result
      strongly depends on the choice of the renormalization and factorization scales used to compute
      NNLO QCD corrections.}
        
    It is also useful to show the results for mixed corrections to the
    production stage only, considering decays of $Z$ bosons in
    the leading-order approximation; this removes the
    dependence of the result on kinematic constraints on the leptons
    that are not well-described in perturbation theory. The
    corresponding results are shown in the fourth column of Table~1.
    It follows from this table that if we consider corrections to the
    production stage {\it only}, the behaviour of individual
    contributions looks better but when corrections are put together,
    mixed NNLO contributions turn out to be only thirty percent smaller than
    the NLO ones. The reason for this seems to be the smallness of the
    NLO corrections, caused by a partial cancellation between QED and
    weak ones inherent to the $G_{\mu}$ scheme, rather than an abnormal
    enhancement of the NNLO mixed QCD-electroweak contributions.

    We turn to the discussion of kinematic distributions. In Fig.~1
    relative corrections to the rapidity and transverse momentum of
    the reconstructed dilepton system are shown for the QCD-QED,
    QCD-weak and QCD-electroweak contributions. Left panes describe
    corrections to the full process that includes production and
    decays of $Z$ bosons; in the right panes we show corrections to
    the production stage only. NNLO QCD corrections rescaled by a factor
    $1/10$ are also shown there, to put the relevance of other
    contributions into perspective.  Similar to the inclusive case,
    we observe that {\it weak
      corrections} are often not negligible when compared to QED
    corrections and, in case of production, they are actually the
    dominant ones.  At the same time, we also observe that the relative
    importance of NNLO QCD and mixed corrections depends on the observable
    and kinematic range. For example, in the central rapidity region NNLO QCD corrections are somewhat smaller than the mixed ones
    but the situation becomes opposite at large rapidities. Similarly, NNLO QCD corrections at large $p_{t,ll}$ are dominant
    whereas at smaller values of the transverse momenta NNLO QCD and mixed QCD-electroweak contributions may be comparable.

    In Fig.~2 we show two distributions that depend on kinematic
    features of individual leptons. In the upper panes, we present the
    transverse momentum distribution of the hardest lepton; in the two
    lower ones we show the distribution in the Collins-Soper angle
    $\theta^*$~\cite{cs}, in the rapidity window $0.6<|y_{ll}|<1.2$.
    This angle can be computed from lepton momenta in the laboratory frame
    using the following formula
    \be
    \cos \theta^* =
    \frac{{\rm sgn}(p_{z,l^+ l^-}) (P_{l^-}^+ P_{l^+}^- - P_{l^-}^-
      P_{l^+}^+)}{\sqrt{m_{l^+l^-}^2 \left ( m_{l^+ l^-}^2 + p_{t,
          l^+ l^-}^2 \right ) }},
    \ee
    where $P_i^\pm = E_i \pm
    p_{i,z}$.
    Studies of the $\cos \theta^*$ distribution at the LHC allow
    for a precise determination of the weak mixing angle.

    The major features of distributions shown in Fig.~2 are similar to
    what we have seen already in Table~1 and Fig.~1. When corrections
    to production and decay are included, mixed QCD-QED corrections
    play an important, sometimes the dominant role; when only corrections
    to the production stage are considered, weak effects become more pronounced
    than QED ones. In the case of the $\cos\theta^*$ distribution, weak and
    QED corrections have similar magnitude even in the case when full
    corrections to the $pp \to Z \to l^+l^-$ process are considered. As 
    is well-known, the
    spikes in corrections to $p_{t,l}$ distributions
    are caused by an
    interplay of cuts on lepton momenta and the leading-order kinematic
    boundary $p_{t,l} < M_Z/2$. Not surprisingly, they are much
    more pronounced when QED corrections to decays are included.

    {\bf Conclusions} We have presented the first complete computation
    of mixed QCD-electroweak corrections to the production of on-shell
    $Z$ bosons in hadron collisions and their subsequent decay to a
    pair of massless electrons.  We find that 
    mixed corrections are about a few permille.
    The only exceptions are QCD-QED corrections to the inclusive process
    and QCD-QED corrections to the production stage -- both 
    at the inclusive level and in the fiducial region --  which are smaller.
    However, corrections strongly
    depend on the imposed kinematic constraints and, in general, do
    not follow a clear hierarchy that would allow an 
    approximate but reliable treatment of them. As we mentioned in the
    introduction, given the smallness of these mixed corrections, the
    $Z$ boson case is, perhaps, not very interesting
    phenomenologically. However, an ambitious goal of extracting the
    mass of the $W$ boson from the LHC data with very high precision calls
    for a complete computation of mixed QCD-electroweak correction to
    the $W$ production process.  We look forward to this interesting
    challenge.

    {\bf Acknowledgments}
    We would like to thank R.~Bonciani, N.~Rana and A.~Vicini for useful
    discussions. R.R. is grateful to the Oxford Physics Department for
    hospitality extended to him during work on this paper. 
    This research is partially supported by BMBF
grant 05H18VKCC1 and by the Deutsche Forschungsgemeinschaft (DFG,
German Research Foundation) under grant 396021762 - TRR 257. The
research of F.B. and F.C. was partially supported by the ERC Starting
Grant 804394 \textsc{hipQCD}.

\end{document}